\documentclass[12pt]{iopart}
\usepackage{graphicx}
\usepackage[bookmarks,dvips,pdfhighlight=/O,pdfstartview=FitH]{hyperref}

\usepackage{iopams}  

\newcommand{\beq}{\begin{equation}}
\newcommand{\eeq}{\end{equation}}
\newcommand{\bea}{\begin{eqnarray}}
\newcommand{\eea}{\end{eqnarray}}
\newcommand{\bce}{\begin{center}}
\newcommand{\ece}{\end{center}}

\newcommand{\dd}{\mathrm{d}}
\def\lsim{\mathrel{\rlap{\lower4pt\hbox{\hskip1pt$\sim$}}
    \raise1pt\hbox{$<$}}}         
\def\gsim{\mathrel{\rlap{\lower4pt\hbox{\hskip1pt$\sim$}}
    \raise1pt\hbox{$>$}}}         

\begin{document}

\title{Heavy-Quark Kinetics in the QGP at LHC}

\author{H van Hees$^1$, V Greco$^2$, R Rapp$^1$} 

\address{$^1$Cyclotron Institute and Physics Department, 
       Texas A\&M University, College Station, Texas 77843-3366, U.S.A.}
\address{$^2$Dipartimento di Fisica e Astronomia, Via S. Sofia 64,
  I-95125 Catania, Italy}

\date{June 29, 2007}

\begin{abstract}
  We present predictions for the nuclear modification factor and
  elliptic flow of $D$ and $B$ mesons, as well as of their decay
  electrons, in semicentral Pb-Pb collisions at the LHC.  Heavy quarks
  are propagated in a Quark-Gluon Plasma using a relativistic Langevin
  simulation with drag and diffusion coefficients from elastic
  interactions with light anti-/quarks and gluons, including
  non-perturbative resonance scattering.  Hadronization at $T_c$ is
  performed within a combined coalescence-fragmentation scheme.
\end{abstract}

In Au-Au collisions at the Relativistic Heavy Ion Collider (RHIC) a
surprisingly large suppression and elliptic flow of ``non-photonic''
single electrons ($e^\pm$, originating from semileptonic decays of $D$
and $B$ mesons) has been found, indicating a strong coupling of charm
($c$) and bottom ($b$) quarks in the Quark-Gluon Plasma (QGP).

We employ a Fokker-Planck approach to evaluate drag and diffusion
coefficients for $c$ and $b$ quarks in the QGP based on elastic
scattering with light quarks and antiquarks via $D$- and $B$-meson
resonances (supplemented by perturbative interactions in color
non-singlet channels)~\cite{vanHees:2004gq}. This picture is motivated
by lattice QCD computations which suggest a survival of mesonic states
above the critical temperature, $T_c$.  Heavy-quark (HQ) kinetics in the
QGP is simulated with a relativistic Langevin
process~\cite{vanHees:2005wb}.  Since the initial temperatures at the
LHC are expected to exceed the resonance dissociation temperatures, we
implement a ``melting'' of $D$- and $B$-mesons above
$T_{\mathrm{diss}}$=$2 T_c$=360~MeV by a factor
$(1+\exp[(T-T_{\mathrm{diss}})/\Delta])^{-1}$ ($\Delta$=50~MeV) in the
transport coefficients.

The medium in a heavy-ion reaction is modeled by a spatially homogeneous
elliptic thermal fireball which expands isentropically.  The temperature
is inferred from an ideal gas QGP equation of state with $N_f$=2.5
massless quark flavors, with the total entropy fixed by the number of
charged hadrons which we extrapolate to $\dd N_{\mathrm{ch}}/\dd
y$$\simeq$1400 for central $\sqrt{s_{NN}}$=5.5~TeV Pb-Pb collisions.
The expansion parameters are adjusted to hydrodynamic simulations,
resulting in a total lifetime of $\tau_{\mathrm{fb}}$$\simeq$6~fm/c at
the end of a hadron-gas QGP mixed phase and an inclusive light-quark
elliptic flow of $\langle v_2 \rangle$=7.5\%.  The QGP formation time,
$\tau_0$, is estimated using of $\tau_0 T_0$=const ($T_0$: initial
temperature), which for semicentral collisions (impact parameter
$b$$\simeq$7~fm) yields $T_0$$\simeq$520~MeV.

Initial HQ $p_T$ spectra are computed using PYTHIA with parameters 
as used by the ALICE Collaboration. $c$ and $b$ quarks are hadronized
into $D$ and $B$ mesons at $T_c$ by coalescence with light 
quarks~\cite{Greco:2003vf}; ``left over'' heavy quarks are 
hadronized with $\delta$-function fragmentation. For semileptonic 
electron decays we assume 3-body kinematics~\cite{vanHees:2005wb}. 
\begin{figure}[t]
\begin{center}
\begin{minipage}{0.345\textwidth}
\includegraphics[width=\textwidth]{quark-minbias-RAA-lhc.eps}
\end{minipage}\hspace*{0.7cm}
\begin{minipage}{0.345\textwidth}
\includegraphics[width=\textwidth]{quark-minbias-lhc-v2.eps}
\end{minipage}

\begin{minipage}{0.355\textwidth}
\includegraphics[width=\textwidth]{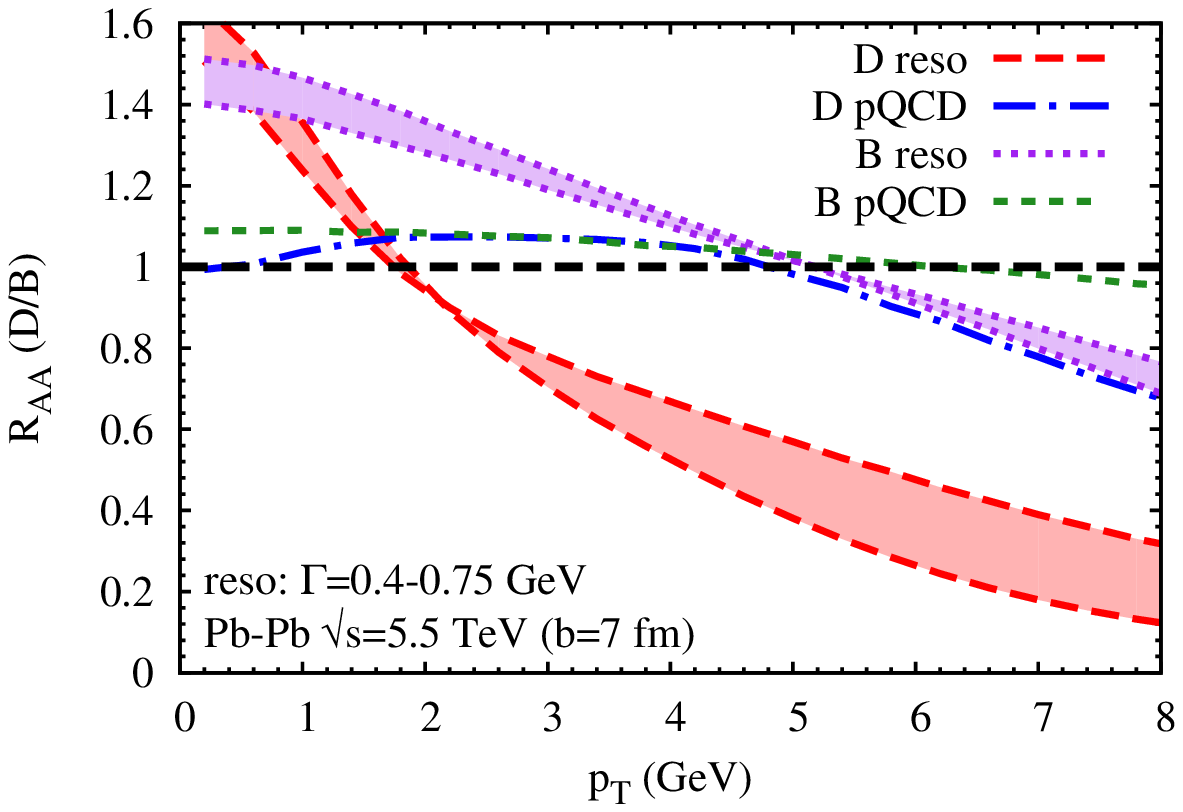}
\end{minipage}\hspace*{0.7cm}
\begin{minipage}{0.355\textwidth}
\includegraphics[width=\textwidth]{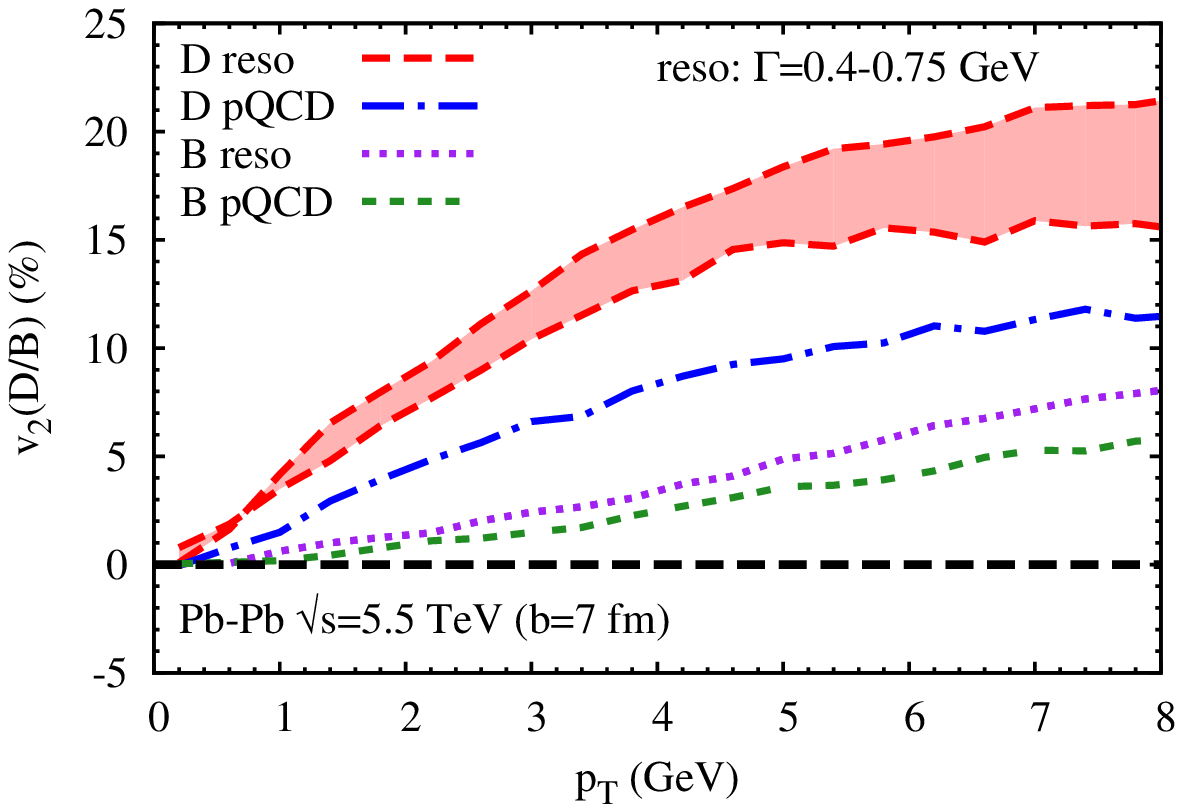}
\end{minipage}

\begin{minipage}{0.355\textwidth}
\includegraphics[width=\textwidth]{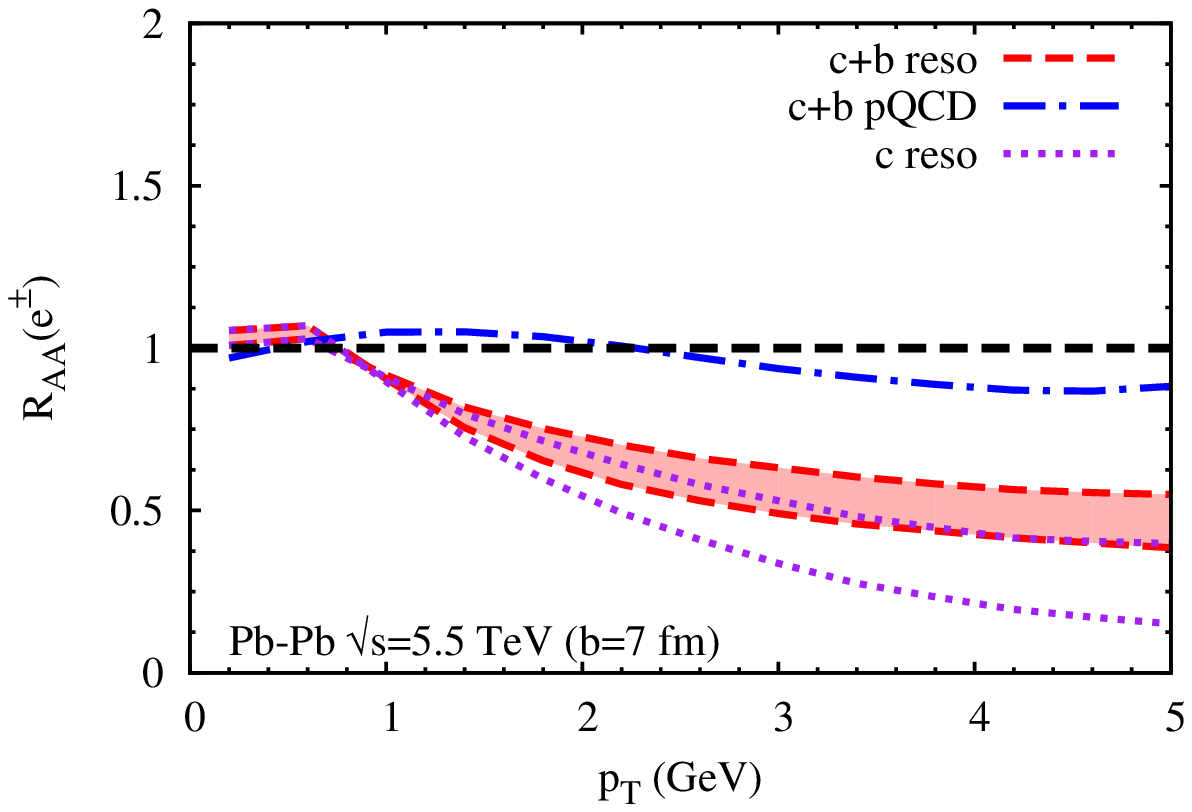}
\end{minipage}\hspace*{0.7cm}
\begin{minipage}{0.355\textwidth}
\includegraphics[width=\textwidth]{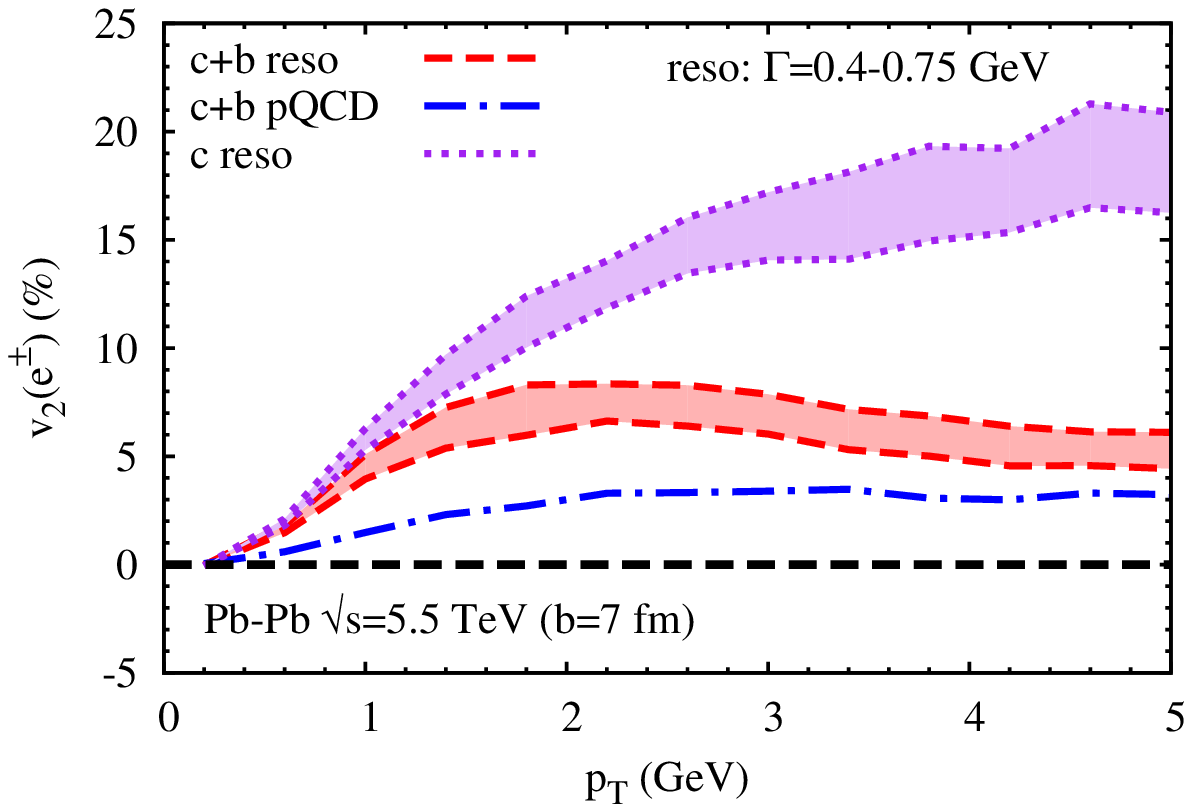}
\end{minipage}

\end{center}
\vspace*{-0.5cm}

\caption{(Color online) Predictions of relativistic Langevin simulations
  for heavy quarks in a sQGP for $b$=7~fm $\sqrt{s_{NN}}$=5.5~TeV Pb-Pb
  collisions: $R_{AA}$ (left column) and $v_2$ (right column) for heavy
  quarks (1$^\mathrm{st}$ row), $D$ and $B$ mesons (2$^\mathrm{nd}$ row)
  and decay-$e^\pm$ (3$^\mathrm{rd}$ row).}
\label{fig.1}
\end{figure}

Fig.~\ref{fig.1} summarizes our results for HQ diffusion in a QGP in
terms of $R_{AA}(p_T)$ and $v_2(p_T)$ at the quark, meson and
$e^\pm$ level for $b$=7~fm Pb-Pb collisions at the LHC
(approximately representing minimum-bias conditions). Our most
important findings are: (a) resonance interactions substantially
increase (decrease) $v_2$ ($R_{AA}$) compared to perturbative
interactions; (b) $b$ quarks are much less affected than $c$ quarks,
reducing the effects in the $e^\pm$ spectra; (c) there is a strong
correlation between a large $v_2$ and a small $R_{AA}$ at the quark
level, which, however, is partially reversed by coalescence
contributions which increase \emph{both} $v_2$ and $R_{AA}$ at the meson
(and $e^\pm$) level. This feature turned out to be important in the
prediction of $e^\pm$ spectra at RHIC; (d) the predictions for LHC are
quantitatively rather similar to our RHIC
results~\cite{vanHees:2005wb,Rapp:2006ta}, due to a combination
of harder initial HQ-$p_T$ spectra with a moderate increase in
interaction strength in the early phases where non-perturbative
resonance scattering is inoperative. 

{\it This work is supported by a U.S. NSF CAREER Award, grant no. PHY-0449489.}





\end{document}